\documentclass {aa}
\usepackage{epsfig}
\usepackage{natbib}    
\usepackage{varioref}   
\usepackage{journals} 
\begin{document}

\thesaurus{  }
 
\title{A continuous low star formation rate in IZw~18 ? \thanks{This research has made use of NASA's Astrophysics Data System Abstract Service.}}

\author{F. Legrand \inst{1,5}, D. Kunth \inst{1} , J.-R. Roy \inst{2}, 
J.M. Mas-Hesse \inst{3} \thanks{JMMH supported by Spanish CICYT under
  grant ESP95-0389-C02-02.}, and J.R. Walsh~\inst{4}~\thanks{DK and JRR
Visiting 
astronomers at Canada-France-Hawaii Telescope, which is operated by the 
National Research Council of Canada, the Centre National de la Recherche 
Scientifique de France, and the University of Hawaii} }
\offprints{F. Legrand, legrand@iap.fr}

\institute{
  Institut d'Astrophysique de Paris, CNRS, 98bis boulevard Arago, F-75014 
  Paris, France.
  \and
  D\'epartement de physique and Observatoire
      du mont M\'egantic, Universit\'e Laval, Qu\'ebec
      Qc G1K 7P4
  \and
  LAEFF-INTA, Apdo 50727, E-28080 Madrid,
  Spain.
  \and
  European Southern Observatory, Karl-Schwarzschild-Str. 2, 
  D-85748 Garching, 
  Germany 
  \and 
  Instituto Nacional de Astrofisica, Optica y Electronica,
  Tonantzintla, Apartado Postal 51 y 216, 72000 Puebla, MEXICO
}

\date{received 08/09/1999 ; accepted 17/01/2000}

\maketitle
\markboth{ Legrand et al.: A continuous low star formation rate in
  IZw~18 ?}{  }
\begin{abstract}

Deep long-slit spectroscopic observations of the blue compact galaxy
IZw~18 obtained with the CFH 3.6 m Telescope are presented. The very
low value of oxygen abundance previously reported is confirmed and a
very homogeneous abundance distribution is found (no
variation larger than 0.05 dex) over the whole ionized region. We
concur with \cite{TT96} and \cite{DRD97} that the observed abundance
level cannot result from the material ejected by the stars formed in
the current burst, and propose that the observed metals were formed
in a previous star formation episode.  Metals ejected in the current
burst of star formation remain most probably hidden in a hot phase and
are undetectable using optical spectroscopy.  We discuss different
scenarios of star formation in IZw~18. Combining various observational
facts, for instance the faint star formation rate observed in low
surface brightness galaxies \citep{VZHSB97}, it is proposed that a low and
continuous rate of star formation occurring during quiescent phases
between bursts could be a significant source of metal enrichment of
the interstellar medium.

  \keywords{Galaxies --
            Galaxies:ISM --
            Galaxies: enrichment of ISM --
            Galaxies:  --
            Galaxies: IZw18}
\end{abstract}
 

\section{ Introduction }
The blue compact galaxy IZw~18 is a fiducial object - it has a very
low metallicity and is currently experiencing an intense star
formation episode.  Its metallicity \citep{SS72,SK93} is the lowest
(1/50 Z$_{\odot}$) observed in this kind of objects and more
metal-deficient blue compact dwarf galaxies have not been found
despite extensive searches \citep{T82,TMMMC91,MMCA94,ITL94,TST96}, with the
possible exception of SBSG 0335-052W \citep{LCIFKH99}. This led
\cite{KS86} to suggest that during a single starburst event, the
metals ejected by the massive ionizing stars are mixed within a short
time scale in the HII region and lead, in few Myr, to a metallicity
level comparable to that of IZw~18. Other studies have also shown that
only one burst is sufficient to account for the oxygen abundance in
IZw~18 \citep{ABP78,LMJDK81,KMM95}. Thus if the metallicity measured
in IZw~18 is solely the result of the metals produced in the current
burst, a discontinuity in the spatial abundance distribution would be
expected corresponding to the edge of the recently enriched region,
typically a few hundred parsecs away from the young stellar core
\citep{RK95}.

Early measurements of the abundance in the HI halo of IZw~18 by
\cite{KLSV94} indicated that the metallicity of the very massive
neutral cloud in front of the ionizing cluster might be a factor of
about 20 lower than in the HII region, strengthening the possibility
of a sharp abundance drop.  However the UV absorption lines they used
were saturated, and this result remains very uncertain
\citep{PL95}. Moreover, recent HI observations by \cite{VZWH98}
displayed lower velocity dispersion in the HI halo than those assumed
by \cite{KLSV94}, leading to a metallicity comparable to the abundance
in the central ionized region.

On the other hand, several observations of starburst galaxies \cite[][
and references therein]{KS97b} have shown no significant gradient or
discontinuity in the oxygen abundance distributions within the HII
regions, except for the well established local overabundance of
nitrogen in NGC5253 \citep{W70,WR87,WR89, KSRRW96,KSRWR97}. This
corroborates models which predict that during a starburst, the heavy
elements produced by the massive stars are ejected with high
velocities into a hot phase, leaving the starburst region without
immediate contribution to the enrichment of the insterstellar medium
\citep{PC87,TT96,DRD97,KS97b,PIL99}. In this scenario, the metals
observed now would have their origin in a previous star formation
event, and an underlying old stellar population would be
expected. Early observations of IZw~18 did not reveal clearly such an
old population \citep{T83,HT95}, but recent reanalysis of HST archive
data \citep{ATG99} has shown that stars older than 1 Gyr must be
present. Moreover, \cite{O99} studied the resolved stellar
population in the near 
infrared with NICMOS onboard HTS and found also that while the NIR colour
magnitude diagram was dominated by stars 10-20 Myr old, numerous red AGB
stars require a much higher age, in agreement with \cite{ATG99}.
NICMOS data require stars older than 1 Gyr to be present and an age as
high as 5 Gyr is favoured. This holds even if a distance slightly higher 
than the conventional 10 Mpc is adopted.
This suggests that the present star formation episode in
IZw~18 is not the first one. The rather high C/O ratio observed in
IZw~18 \citep{GSDS97} could also suggest a carbon enrichment by an
evolved population of intermediate mass stars. However, other
starburst galaxies show quite lower C/O ratio
\citep{GSDPTPTTS95,KS98,IT99}, so this fact remains puzzling and
controversial \citep{IT99}. Considering the large uncertainties on the
determinations of the stellar yields \citep{PR98,P99} and on the
determination of the C/O ratio \citep{IFGGT97,IT99}, this may not be
used as a strong evidence for an enrichment by an old stellar
population.

Thus the mechanism responsible for the dispersal and mixing of newly
synthesized elements in a starburst galaxy remains unclear, as well as
the chain of star formation events responsible for the observed
abundances. IZw~18, as the lowest abundance galaxy among starbursts,
is an ideal laboratory to study these processes; its low metallicity
is indicative of a rather ``simple'' star formation history, and one
would expect the material ejected by the present massive stars to give
high contrast in the abundances between the enriched and the non
enriched zones. However, if the small companion galaxy northwest of I
Zw18 has had an influence, as suggested by \cite{DEC96} through tidal
effects or streaming gas resulting from a collision with the main body
of I Zw18, the recent history of dispersal and mixing of elements may
not be that easy to disentangle.

We conducted deep long slit spectroscopy of IZw~18 in order to measure
the O/H abundances as far as possible from the central HII region of
the NW knot, and to detect a discontinuity or systematic gradient in
the metallicity distribution. Observations and reduction are described
in section \ref{section:observations}; results are presented in
section \ref{section:extinction} and \ref{section:abundances} and the
star formation history of this galaxy is discussed in the last
sections.


\section{ Observations and data reduction } \label{section:observations}

\begin{figure}
\psfig{figure=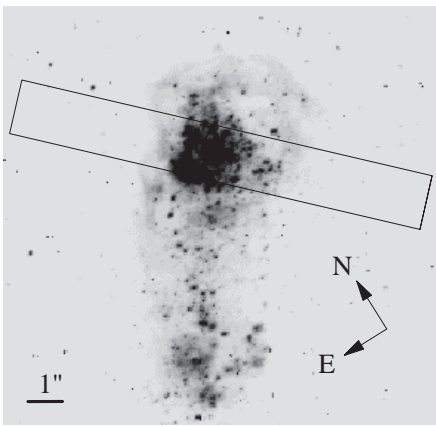,clip=,bbllx=30pt,bblly=30pt,bburx=335pt,bbury=325pt,height=6cm,angle=0}
\caption[]{V-band image of IZw~18 \citep[from ][]{HT95} with the slit position
  overlaid.} \label{fig:slitpos}
\end{figure}

Seventeen exposures of 3000 seconds each of the blue compact galaxy
IZw~18 were obtained with the 3.6 m Canada-France-Hawaii Telescope
during three successive nights between 1995 February 1 and 4 using the
MOS spectrograph with the 2048 $\times$ 2088 Loral 3 thick CCD
detector.  A long slit (1.52 arcsec wide) was used with a position
angle of 45 $\degr$ covering the spectral range from 3700 to 6900 \AA
.  The position of the slit is displayed on Fig~\ref{fig:slitpos}.
The spatial resolution was 0.3145~arcsec/pix and the dispersion
1.58~\AA/pix, leading to a spectral resolution of about 8.2~\AA.  The
seeing was between 1 and 1.5 arcsec.  The spectra were reduced using
IRAF. The bias was removed using the overscan section from each
frame. The pixel-to-pixel sensitivity correction and the illumination
effects (vignetting) were corrected using dome flat field and sky flat
images.  The images were calibrated in wavelength using a combination
of two exposures made during the second night with a Neon and a Helium
lamp respectively.  Five 50~seconds exposure images of the standard
star Feige~34 were obtained in order to flux calibrate the spectra.
To account for wavelength-dependent atmospheric refraction, we fitted
a low-order polynomial along the stellar continuum in each frame and
then realigned each spectrum before combination.  The $H\beta$ spatial
profile on the seven frames obtained during the first night was
different from those ten obtained during the two other nights. We
assumed that the positioning of the slit was slightly different during
the first and the two other nights. Nevertheless, as the offset was
less than 1'' (slit positionning error), we aligned and combined all
the nights together in order to increase the S/N ratio. \\ After
reduction, an abnormal ``diffuse light'' background in the blue part
of the long exposure images appeared. The origin of this ``light'' is
probably due to a slight increase in the temperature of the CCD with
time or to light diffused in the instrument during long exposures.
This feature was removed by the subtraction of the background (task
BACKGROUND) and using the task APSCATTER which is especially designed
for this kind of purpose. The residuals after correction were less
than 0.5 $\%$ of the continuum level. Three bad columns (579 to 581
{\it i.e.\/,} 4600 and 4602~\AA\ respectively) of the CCD were
ignored. We applied a Doppler correction to shift the final spectrum
to zero velocity. \\

\begin{figure*}
\psfig{figure=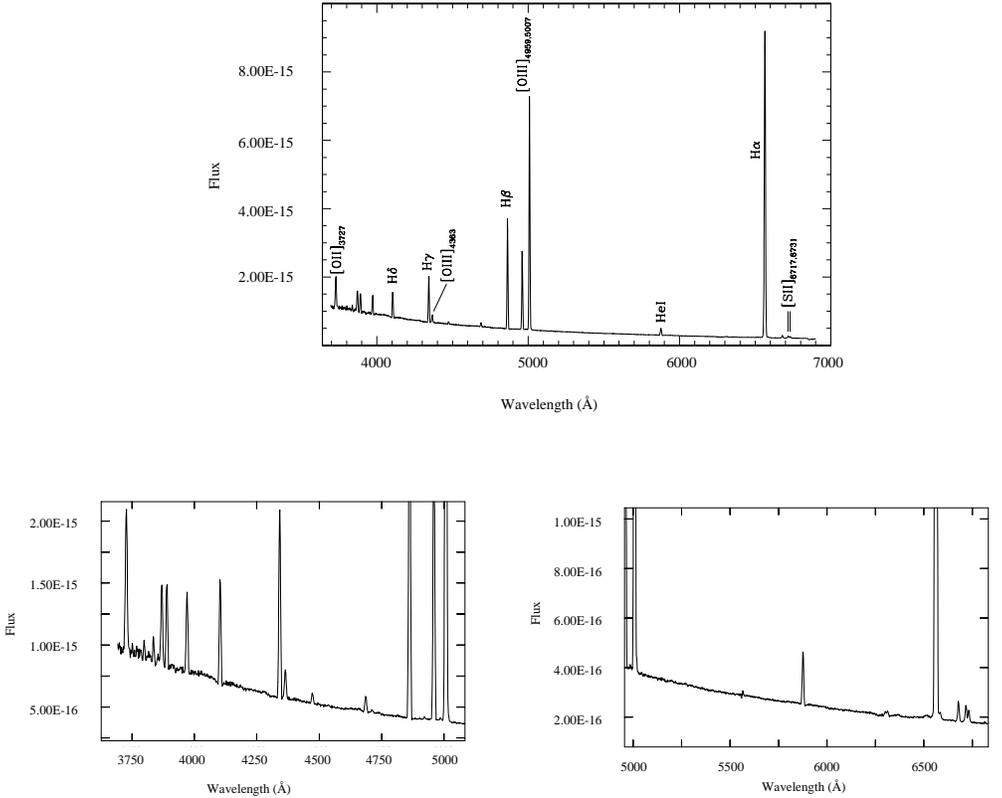 ,clip=,bbllx=0pt,bblly=0pt,bburx=650pt,bbury=530pt,height=11cm,angle=0}
\caption[]{Large aperture spectrum (over 25 pix {\it i.e.\/,} 7.85'') of
  IZw~18, with zooms on the blue and red part to show the faintest
  lines. The most important lines are labelled. } \label{fig:spectinteg}
\end{figure*}

Spectra were extracted by summing along the slit. The apertures used
were 5 pixels (1.57'') wide with 2 pixels (0.63'') of overlap. In
order to increase the signal to noise (especially for the [OIII]4363\AA\
line), large aperture spectrum were extracted summing over 12 pixels
(3.78'') every 6 pixels (1.89'') along the slit, but this did not
allow extension of the region over which [OIII]4363\AA\ could be measured.
We also extracted a large aperture spectrum integrated over the whole
galaxy (25 pixels centered on the maximum of the continuum emission)
in order to compare our observations with the spectroscopic
measurements (but with different PA) of \cite{SK93}. A small aperture
spectrum integrated over 2 pixels (0.62'') has also been extracted to
match the aperture used by \cite{IT99}. The large aperture spectrum is
displayed in figure \ref{fig:spectinteg} and results of line
measurements (for both small and large aperture) are shown in
table~\ref{tab:spectrinteg}. Their mean FWHM is around 8 \AA, and the
lines are unresolved. \\

Emission lines were measured automatically using the routine
TWOFITLINES\footnote{Twofitlines, Version 1.4 package for IRAF
provided by Jose Acosta, Instituto de Astrofisica de Canarias -
SPAIN}.  We compared the measurements
with those made interactively with Gaussian fits
through the IRAF task SPLOT and found no differences larger than two
percent.  A few weak lines in regions of low S/N high, for which no
Gaussian could be fitted, were measured by direct integration.  The
errors bars were computed by summing in quadrature the effective
photon noise on the line flux and the rms noise in the local
continuum. An additional two percent error accounts for
uncertainties in the flat-fielding and sky+diffuse light subtraction
process.


    \begin{table*}
    \caption [] {Observed line fluxes (without reddening correction)
      of the NW component of IZw~18 for the large aperture (LA) and
    the small aperture (SA) spectra.} \label{tab:spectrinteg} 
    \small
    \begin{flushleft}
    \begin{tabular}{cccccccc}
    \hline 
    Line Id & Rest $\lambda$ (\AA) & Flux (LA) & Error on  & EW
    (LA) & Flux (SA) & Error on & EW
    (SA)\\ 
            &                      &           & Flux (LA) &   
         &           & Flux (SA) \\
\hline
$\rm [SII] $    & 6731.4 & 1.6     & 0.1  &  -2.5  & 1.4    & 0.2  & -1.1  \\ 
$\rm [SII] $    & 6716.8 & 2.2     & 0.1  &  -3.5  & 1.8    & 0.2  & -1.5\\ 
$\rm HeI $      & 6678.4 & 2.5     & 0.1  &  -3.9  & 2.7    & 0.2  & -2.1\\ 
$\rm H\alpha$   & 6563.0 & 306.4   & 0.9  &  -450  & 329.8  & 0.2  & -251.8\\ 
$\rm [NII] $    & 6584.2 & 0.8     & 0.1  &  -1.2  & 0.7    & 0.2  & -0.6\\ 
$\rm [SIII]$    & 6312.9 & 0.5     & 0.1  &  -0.7  & detected & ---& --- \\ 
$\rm [OI] $     & 6301.6 & 0.5     & 0.1  &  -0.7  & ---    & ---  & --- \\
$\rm HeI $      & 5876.1 & 6.5     & 0.1  &  -7.3  & 6.3    & 0.2  & -3.6 \\ 
$\rm [OIII]$    & 5007.2 & 205.4   & 0.6  &  -152  & 223.5  & 0.3  & -86.2 \\ 
$\rm [OIII]$    & 4959.3 & 68.8    & 0.6  &  -49.4 & 74.9   & 0.3  & -28.2 \\ 
$\rm HeI $      & 4922.4 & 0.6     & 0.1  &  -0.5  & ---    & ---  & --- \\
$\rm H\beta$    & 4861.9 & 100     & 0.2  &  -68.7 & 100    & 0.3  & -35.8 \\ 
$\rm HeI+[AIV]$ & 4712.0 & 0.9     & 0.1  &  -0.6  & 0.3    & 0.2  & -0.1 \\ 
$\rm HeII $     & 4686.4 & 3.6     & 0.2  &  -2.2  & 4.3    & 0.4  & -1.4 \\ 
$\rm HeI $      & 4472.8 & 2.3     & 0.2  &  -1.2  & 1.2    & 0.5  & -0.3 \\
$\rm [OIII] $   & 4364.4 & 6.6     & 0.3  &  -3.3  & 7.5    & 0.7  & -2.0 \\ 
$\rm H\gamma$   & 4341.8 & 44.8    & 0.2  &  -21.9 & 41.3   & 0.5  & -10.7 \\ 
$\rm H\delta$   & 4103.3 & 22.2    & 0.3  &  -8.9  & 17.0   & 0.7  & -3.6 \\ 
$\rm H\epsilon$ & 3970.7 & 17.6    & 0.5  &  -6.4  & 12.2   & 0.8  & -2.3 \\ 
$\rm [NeIII] $  & 3869.5 & 18.3    & 0.7  &  -6.1  & 19.4   & 1.3  & -3.4 \\ 
$\rm HI+NeIII$  & 3889.9 & 17.4    & 0.6  &  -5.9  & 12.0   & 1.1  & -2.1 \\ 
$\rm SiII $     & 3854.9 & 2.7     & 0.1  &  -0.9  & detected &--- & --- \\ 
$\rm HI $       & 3836.2 & 3.5     & 0.5  &  -1.1  & 1.3    & 0.7  & -0.2 \\ 
$\rm HI $       & 3798.9 & 3.4     & 0.1  &  -1.1  & 1.7    & 0.7  & -0.3 \\ 
$\rm HI $       & 3769.1 & detected& ---  &  ---   & ---    & ---  & --- \\ 
$\rm HI $       & 3751.8 & detected& ---  &  ---   & ---    & ---  & --- \\ 
$\rm [OII] $    & 3727.4 & 35.6    & 0.9  &  -10.9 & 30.1   & 1.8  & -4.8 \\ 
\hline
\hline
$\rm H\beta$flux&        & 2.9E-14 &      &        & 3.5E-15  \\
$\rm (ergs\,cm^{-2}\,s^{-1})$  \\
E(B-V)$_{H_{\alpha}/H_{\beta}}$ &&  0.075  &&& 0.121      \\
E(B-V)$_{H_{\gamma}/H_{\beta}}$ &&  0.008  &&& 0.077      \\
E(B-V)$_{H_{\delta}/H_{\beta}}$ &&  0.008  &&& 0.136      \\
\hline
    \end{tabular}
    \end{flushleft}

    \end{table*}


\section{Dust extinction} \label{section:extinction}

The interstellar dust extinction, or reddening, was first evaluated using
the H$\alpha$/H$\beta$, H$\gamma$/H$\beta$ and H$\delta$/H$\beta$
ratios, assuming their intrinsic values to be 2.75, 0.475, 0.264
respectively for an electron temperature of 20000 K and a density of
100 cm$^{-3}$ \citep{O89}.  We used the extinction function \\ 
\begin{equation}
\rm \frac{I(\lambda)}{I(H\beta)}=\frac{F(\lambda)}{F(H\beta)}10^{K(\lambda).E(B-V)}
\end{equation} 
where I($\lambda$) is the intrinsic line intensity, F($\lambda$)
is the observed flux at each wavelength and K($\lambda$) is the
extinction function according to the galactic reddening law of
\cite{S79}. \\

In order to correct from the effect of stellar absorption in the Balmer
lines, we first assumed that their strength was the same for all the
lines. We derived then the value for which consistent results for E(B-V)
were obtained using the three different Balmer ratios. As the extent over
which the flux was integrated is larger than the size of the ionizing star
cluster, we corrected for the underlying stellar absorption only in the
central area, {\it i.e.\/,} over 3.8'' (185 pc) centered on the maximum
continuum emission, according to the images of \cite{HT95}.  We found the
underlying stellar absorption to be around 1.8 \AA, close to the value of 2
\AA \ used by \cite{SK93} and \cite{RW87} and adopted this value (1.8 \AA)
for correction.  

\begin{figure}
\psfig{figure=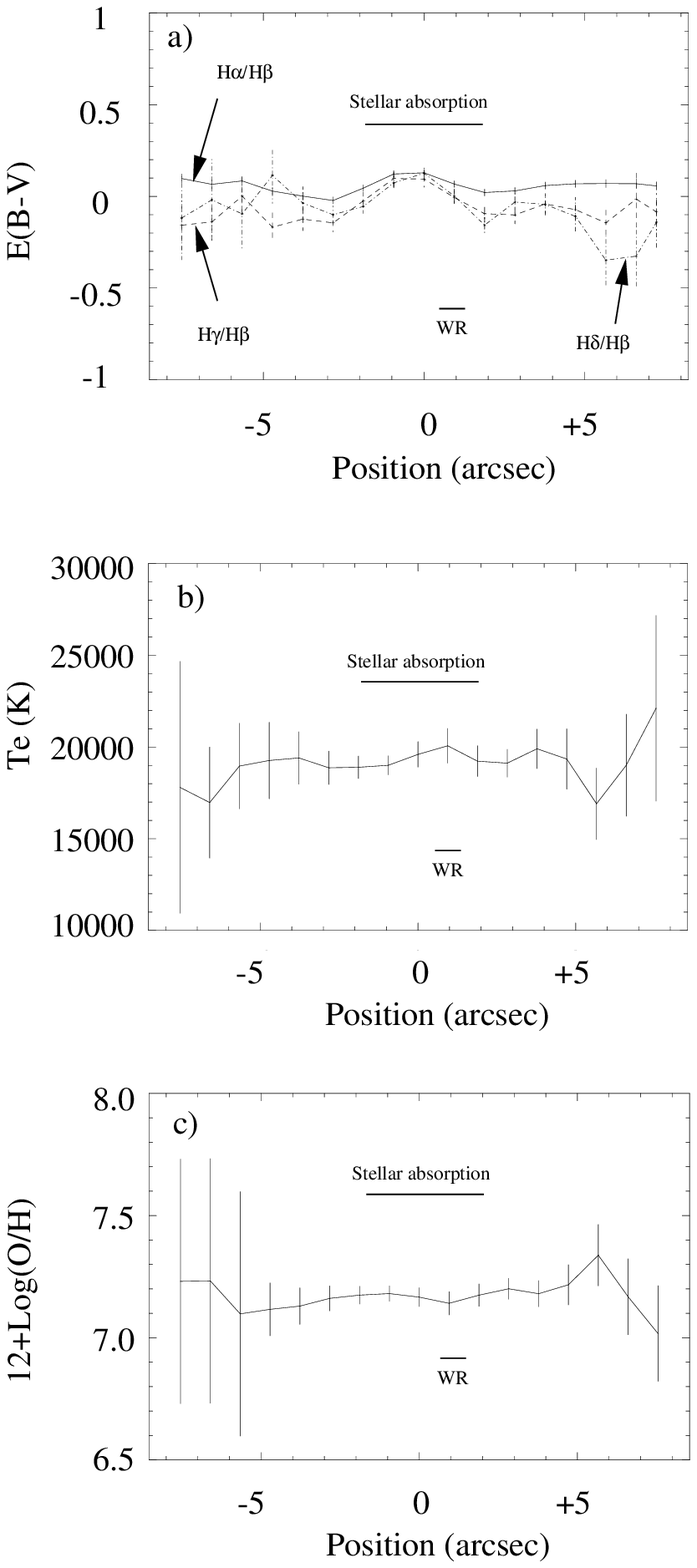,height=15cm,angle=0}
\caption[]{Spatial profile of E(B-V), Te and oxygen abundance in
  IZw~18. Positions are given  in arcsec from the maximum of the
  continuum emission, with values negative and positive in the NE and
  SW directions respectively. The region where the stellar absorption
  has been subtracted is indicated, as well as the region where the WR
  stars have been found.}
  \label{fig:ebmv}
\end{figure}

The variation of the extinction parameter E(B-V) along the slit is shown in
Fig~\ref{fig:ebmv}.a. It can be seen that a good agreement between the
three computed values is obtained only in the central region (we have
indeed forced this agreement by defining the strength of the absorption
lines). Outside the central area, values obtained using H$\gamma$/H$\beta$
and H$\delta$/H$\beta$ ratios are systematically lower than values obtained
with H$\alpha$/H$\beta$, and fall most of the time below zero. Artificially
increasing the H$\beta$ flux by less than four percent erases this
discrepancy, suggesting it could be (partially) related to small
calibration errors dues to the Balmer absorption lines in the Feige 34
spectra. 

\cite{SSC99} have shown that {$\rm
H_{\alpha}$} is partially excited by collisions in IZw~18, so that the
{$\rm H_{\alpha}/H_{\beta}$} ratio should be between 2.95 and 3.00 for at
least the main body of the nebula, higher than for case B
recombination. This effect would explain the discrepancy between the
reddening values estimated using different line ratios. These authors
conclude that the reddening affecting this ionized nebula should be
practically equal to zero. Therefore, we have assumed no reddening at
all along the slit in our calculations.

\section{Abundance determinations} \label{section:abundances}

\subsection{Electron temperature}

The temperature sensitive [OIII]4363\AA\ line was measured over a slit
length of 12 arcsec.  This extent is comparable with that reported by
\cite{M96} despite the large difference in exposure times (51000 s for
this observation and 12000 s for that of Martin), but with a different
orientation of the slit (PA = 45$\degr \ $ against 7.6$\degr \ $ of
Martin).  Nevertheless, the [OIII]5007 line was observed over a length
of 49 arcsec (2.5 kpc) against 23 arcsec in Martin's observations.  We
then computed the ratio of [OIII]4959+[OIII]5007 line strength to
[OIII]4363\AA\ to evaluate the electron temperature $\rm T_{e}$(OIII) with
a program based on the 3-level atom formulae from \cite{MC84} using
atomic data from \cite{M83}.  Uncertainties were propagated through
all steps to derive the error bars.  Fig~\ref{fig:ebmv}.b shows the
variation of the derived electron temperature as a function of
position along the slit. The electron temperature can be considered
constant across the HII region.  Using the large and small aperture
spectra, we also obtain a mean electron temperature of 19300 $\pm$ 600
K and 19700 $\pm$ 1000 K in agreement with the previous determinations
of \cite{SK93} and \cite{IT99} respectively. However, our
measurement in the small aperture appears somewhat smaller (but still
compatible within the error bars) than the value of \cite{IT99} .

\subsection{The oxygen abundance}
 
Oxygen abundances were derived for the regions where the electron
temperature was measured using [OIII]$\lambda4363$.  They were obtained
by summing over ionization states using the expression: \\ 
\begin{equation}
\rm \frac{O}{H} = \frac{O^{+}}{H^{+}} + \frac{O^{++}}{H^{+}} 
\end{equation}
The presence of HeII~4686 suggests that $\rm O^{+++}$ should also be
present.  Generally, HeII~4686 is used to evaluate the abundance of
$\rm \frac{O^{+++}}{H^{+}} $ but the origin, nebular or circumstellar,
of HeII~4686 in IZw~18 is not well established. However it has been
shown \citep{LKRMHW97} that this line peaks at the position of the WR
feature, suggesting that these stars could be responsible for a higher
excitation locally, and that $\rm O^{+++}$ is not an abundant ion.
\cite{SK93} and \cite{IT98} have estimated that this stage contribute
less than 4 percent to the total oxygen abundance.  In term of a
possible abundance gradient in IZw~18, neglecting $\rm O^{+++}$ will
not change the general trend of the abundance profile and would, at
worst, slightly underestimate, by a few percent, the oxygen abundance
in the region around WR stars. So the contribution of $\rm
\frac{O^{+++}}{H^{+}} $ was not included.  The contribution of the
ionization states $\rm O^{+} \ and \ O^{++}$ was computed using a
program based on the 3-level atom formulae from \cite{MC84} with
atomic data from \cite{M83} using the [OII] 3727 and [OIII] 4959
lines.  We compared the abundances delivered by our program with
abundances calculated by IRAF using the 5-level atom approximation
from \cite{SD95} and found no significant differences.  The spatial
profile of the oxygen abundance is given in Fig~\ref{fig:ebmv}.c.

We also used the large and small aperture spectra to derive the mean oxygen
abundance $\rm 12+log(\frac{O}{H})=7.18~\pm~0.03 $ in both cases, in
excellent agreement with \cite{SK93}. This value appears different
from that reported by \cite{IT99}, mainly due to the differences in
the electronic temperatures adopted.

Our results (Fig.~\ref{fig:ebmv}.c) show unambiguously that there is
no significant abundance gradient nor discontinuity in the NW-HII
region of IZw~18 at scales smaller than 600 pc (using \rm $\rm
H_{o}=75 \ km.s^{-1}.Mpc^{-1}$). \cite{M96} suggested a possible, but
weak, gradient in an orthogonal direction. The spatial resolution of
our observations is 50 pc, and smaller scale inhomogeneities cannot be
excluded. Moreover, the spatial profile of the oxygen lines does not
indicate any abrupt change in metallicity at larger distances.
Combined with the results of \cite{VZWH98}, who found  for the HI halo
an abundance comparable with that of the HII gas, our results strongly
favour a homogeneous metallicity distribution over the whole
galaxy. This is fully consistent with related studies which have found
very homogeneous spatial abundance distributions in several other
giant HII regions \citep{DTPVE87,GDPTTVTTTREMHGVDCC94,S85}, as in 30
Doradus \citep{RM87}, LMC-SMC \citep{DH77,PEFW78,RD90}, or dwarf and
irregular galaxies \citep{DRD97,KS97b,KS96,RBDM96,PES80,MMDO91b},
again with the exception of NGC 5253 already refered to in the
Introduction.\\

\section{Toward a new star formation history for
IZw~18}\label{section:guidelines}

\subsection{A previous star formation event}

The oxygen abundance distribution in IZw~18 appears extremely homogeneous
throughout the galaxy, indicating a thoroughly mixed interstellar medium.
If the measured abundances result from the metals ejected by the massive
stars involved in the current burst, as suggested by \cite{KS86}, this
would imply efficient mixing of the ejecta on scales of at least 600 pc
within a timescale comparable to the age of the present burst {\it i.e.\/,}
a few Myr \citep{HT95}. However, dispersal of the heavy elements ejected by
the massive stars can hardly be accomplished in less than $10^{8}$ yr on
scales between 100 and 1000 pc \citep{RK95}; the timescale required for
complete mixing is even longer \citep{TT96}. Thus the observed metals
cannot arise from the material ejected by the stars formed in the current
burst. It follows that {\it the presently observed metals should have been
formed in a previous star formation episode}. The metals ejected in the
current burst of star formation remain most probably hidden in a hot phase
as suggested by \cite{PC87} and more recently by \cite{TT96}, \cite{DRD97},
\cite{KS97b} and \cite{PIL99}. \cite{B99} has shown on a
deep pointing with the ROSAT HRI instrument that there is extended X-ray
emission to the SW and maybe to the NE from the central bubble of 
IZw~18. This extended emission seems to trace the expanding
{$\rm H_{\alpha}$} loops, leading the author to conclude that it supports
the picture of hot, metal-enriched gas streaming out of IZw~18. This gas
would have been ejected into the halo where it would take long excursions
while cooling and returning to the central galactic region to become available
for future processing into stars \citep{TT96}. 
X-ray observations of the BCD VIIZw~403 \citep{PFTL94}  are also
interpreted as hot material ejected by 
the present starburst. The availability of powerful X-ray observatories in
the near future will allow the metallicity of this hot gas to be derived,
allowing this scenario to be confirmed. \\

If the observed metals in IZw~18 were formed in a previous star formation
episode, this would imply that the object is not a ``young'' galaxy
undergoing its first star formation as suggested by \cite{SS72}. Such a
view is also supported by other studies, which independently lead to
another scenario \citep{DGS88,DH90,HT95,KMM95,GSDS96,ATG99}.


\subsection{The dearth of low metallicity galaxies}
\label{subsection:low_metallicity}

The metal abundances measured in IZw~18 are the lowest known in the
interstellar matter (but not in stars) of the local universe; this
remains so despite extensive metallicity measurements in emission line
galaxies \citep{T82,TMMMC91,MMCA94,ITL94,TST96}.  Because of the
correlation between size, luminosity and metallicity in dwarf galaxies
\citep{SKH89}, \cite{MMCA94} proposed that galaxies with very low
metallicity are too faint to be ``caught'' in their sample. This
raises the possibility that extremely metal deficient objects may be
very faint. IZw~18 and other starburst galaxies \citep{RB95} lie quite
far away from the correlation established by \cite{SKH89} for dwarf
irregular galaxies. This may reflect the fact that they are presently
undergoing a strong star formation event which increases their
luminosity. However, the galaxies used by \cite{SKH89} were selected
from the H$\alpha$ catalog of \cite{KEH89}, thus the sample allows for
current star formation! The origin of the correlation remains unclear
\citep[see also][]{S99}.\\

It is easy to show that the present star formation rate in IZw~18 or
in other starbursts cannot be sustained for a Hubble time without
producing excessive chemical enrichment and a numerous stellar
population. It is generally admitted that blue compact galaxies
experience violent star formation events separated by long quiescent
phases \citep{SS72} during which they would appear as Low Surface
Brightness Galaxies (LSBG) or
quiescent dwarfs. However this population does not contain any objects
more metal poor than IZw~18 \citep{MGB93,MG94,RB95,VZHS97a,VZHS97b}. 
{\it Does the metallicity of IZw~18 represent a lower limit for the
abundance in the gas of local galaxies?} If so, why?

\subsection{The lack of HI clouds without optical counterpart}

Different observing programs have been carried out to search for
isolated intergalactic HI cloud, but without success so far
\citep{B97}. Most local so-called primeval HI clouds candidates turned
out to be associated with stars \citep[see for example][ for HI1225+01]
{DJ90,IBMSS90,MCMIGHWH90,SDSAMGH91,CGH95}. {\it Does this mean that
such entities do not exist?} If so, this would imply that {\it all
gas clouds (with a mass comparable to that of a dwarf galaxy) have
formed stars}. However, the detection limits for HI surveys remain
quite high ($N_{HI} \sim 10^{18} \ \rm cm^{-2}$), and the existence of
very small primeval HI clouds cannot be ruled out. Nevertheless, if
isolated dwarf galaxy progenitor HI clouds existed, they would have
sizes and masses comparable to small galaxies, and they would present
sufficient column densities to be detected by radio techniques. So
far, non-detection indicates that if such clouds exist, they are very
rare. This idea is reinforced by the presence of absorption line
systems of high column densities in the spectra of quasars which seems
to arise mainly from halos of bright galaxies and not from small HI
clouds \citep{LBTW95,TLS97}, indicating again that the latter are
sparse.  Furthermore, it has been shown that the diffuse cosmic UV
background can ionize the extreme outer HI disks of spiral galaxies
\citep{VG91,CSS89,MA90,CS93a,CS93b}, producing an abrupt fall in their
HI column density. This effect could contribute to hide some primeval
HI gas from the current surveys.

\subsection{The temporal evolution of the metallicity}

\begin{figure}
\psfig{figure=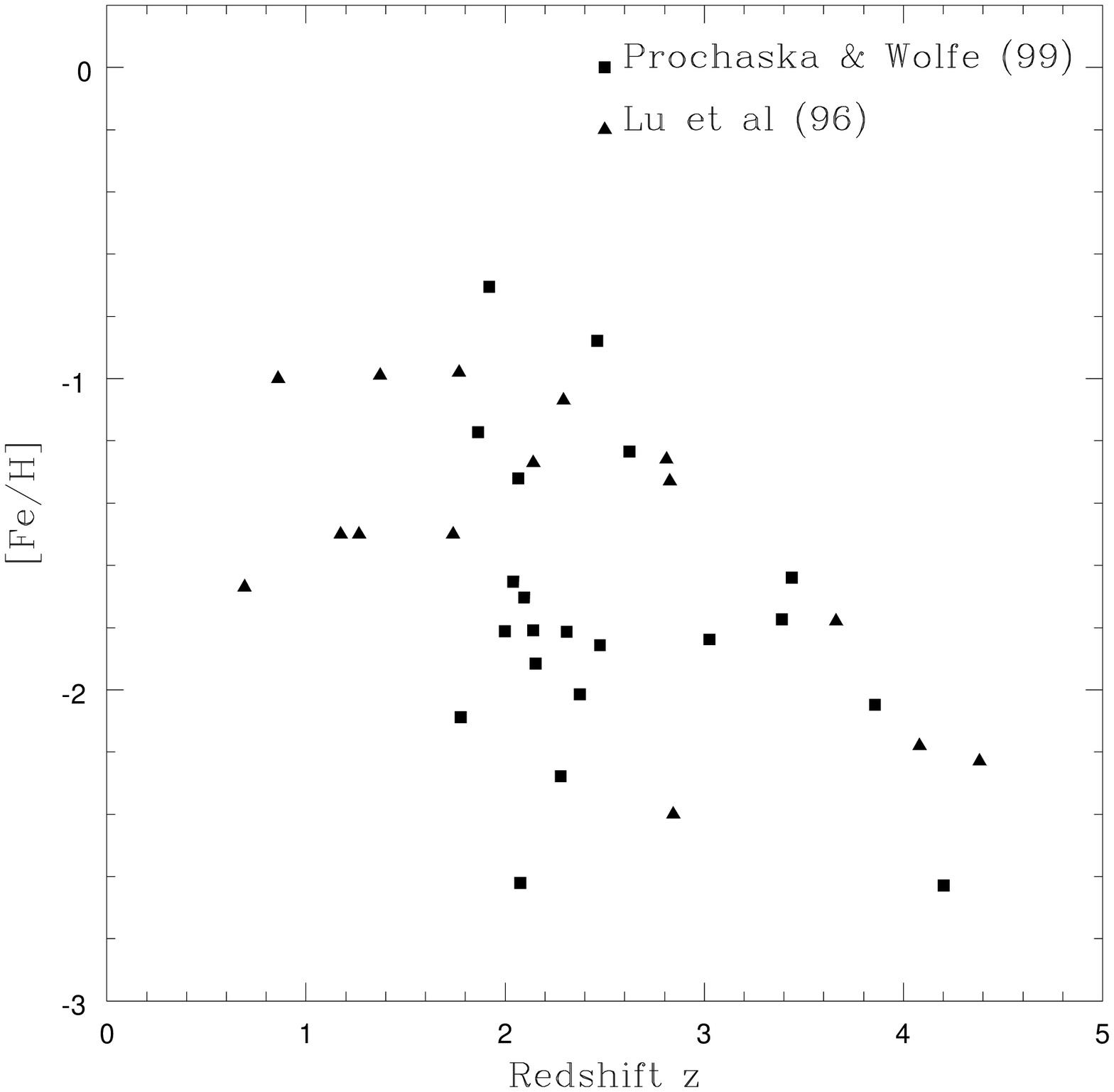,height=8cm,angle=0}
\caption[]{DLA [Fe/H] abundance as a function of redshift. Data from
  \cite{LSBCV96} and \cite{PW99}}
  \label{fig:fedla} 
\end{figure}

Absorption lines in Damped Lyman Alpha (DLA) systems are used to study
the temporal evolution of the metallicity of the interstellar
gas. Although the nature of the absorbing systems is still
controversial \citep{TLS97}, it is generally admitted that the
metallic lines are associated in some way with galaxies. The temporal
evolution of the metallicity in the DLA systems reported by
\cite{LSBCV96} and more recently by \cite{LSB98} and \cite{PW99} is
reproduced in Fig~\ref{fig:fedla}.

One notices that the mean metallicity of the interstellar gas
increases as one gets closer to local time.  This is generally
interpreted as the effect of cumulative enrichment by strong star
formation events.  However, a more intriguing feature is that the
metallicity of the most underabundant systems seems also to increase
with time! No extremely underabundant system has been found at low
redshift. If the enrichment is solely the result of starburst events,
we should find, locally, objects which have not undergone any burst
(or very few of them) ; these objects would have a very low
metallicity (comparable to what is observed at high redshift). {\it
Does the apparent increase in metallicity of the most underabundant
systems indicate the existence of a minimal and continuous enrichment
of the interstellar medium?} The number of systems observed at low
redshift is small \citep{MY92,SBBD95,PB97,DLVR97,BLBBD98,SPSGVGLC98};
if some unevolved systems exist, they must be very few.  The non
detection of such systems could arise from a selection effect rather
than from their inexistence.

\subsection{A new star formation regime}

It is generally accepted that the metal enrichment of the ISM builds
up mainly in bursts.  Different studies have been carried out to model
these bursts to reproduce the global properties of galaxies.  In the
case of IZw~18, \cite{KMM95} have shown that only one burst, with an
intensity comparable to the present one, is enough to produce the
observed abundances. As we have shown, this single burst cannot be the
present one. Previous massive star formation has occurred. We cannot
eliminate the possibility that this previous star formation event was
a starburst.

However, starburst episodes must be separated by quiescent phases,
during which these systems appear as quiescent dwarfs or Low Surface
Brightness Galaxies (LSBG). Studies of the latter objects
\citep{VZHSB97} have revealed that, despite their low gas density,
star formation occurs (with a weak efficiency), probably as a local
process instead of a global event. {\it The SFR between bursts is very
low, but not zero, so the metallicity would increase slowly during
these quiescent phases.}  Because these star formation rates are very
weak, they are generally neglected in studies of star formation
history of galaxies. However, in dealing with very low metallicity
galaxies, they are capable of raising the metallicity levels up to
values comparable to that of IZw~18 in less than a Hubble time.

For example the galaxy UGC~9128, studied by \cite{VZHS97b} presents a
SFR of about $1.7\,10^{-4}\ M_{\odot}\,yr^{-1}$ for a HI mass of
$3.55\,10^{7}\ M_{\odot}$. If such a low SFR lasts 10 Gyr, it will
form $1.7\,10^{6}\ M_{\odot}$ in stars, and no more than 5\% of the
initial mass of gas will have been transformed into stars.  At this
low continuous star formation rate, sustained during even a Hubble
time, the fraction of gas still available at present epoch remains
high (about 95\%). Thus the existence of a continuous low star
formation rate in dwarf galaxies is consistent with the large HI
reservoirs generally observed in these objects.

The current metallicity of the gas $Z_{gas}$ assuming the simple
closed box model \citep{P98} can be expressed by \citep{SS72}

\begin{equation}
Z_{gas}\,\sim\,-y\,Ln(G)
\end{equation}

\noindent where $G$ is the fraction of gas presently available and $y$
the yield in heavy elements. The uncertainties on this last parameter
are large, but $y$ is likely to be in the range 0.01 to 0.036
\citep{M92}.  Using a mean value $y \sim 0.02$, and for the example
above with $G=0.95$, we estimate the metallicity of the gas resulting
from this low SFR enrichment to be close to $10^{-3}$, that is 1/20th
solar!

Consequently, a low continuous star formation rate cannot be neglected, 
especially when dealing with low
metallicity galaxies; this may be the dominant star forming
and metal enrichment process in dwarf galaxies.  

We propose that in the most extreme objects, like IZw~18, a continuous
low star formation regime can account for the observed abundances.  We
surmise that the present starburst is the first major one in the
history of IZw~18, and that a mild star formation rate has been going
on for several Gyr.  Preliminary calculations strengthen this
hypothesis \citep{LK98}. If such a low regime is universal, we expect
that all small systems have been forming stars and that their
metallicity has increased slowly but steadily with time.  This
scenario explains the lack of local objects more underabundant than
IZw~18, the absence of HI clouds without an optical counterpart and
the evolution as a function of redshift of the most metal poor quasars
absorption systems.  Detailed modelling of this low star formation
regime is presented in \cite{L99}.


\section{Conclusion}

We have acquired deep long slit spectroscopy of the metal poor dwarf
star forming galaxy IZw 18. We confirm the very low metal content of
the galaxy, and show that no significant abundance gradient nor
inhomogeneities larger than $\pm 0.05$ dex are present in IZw 18 on
scales of 50 pc to 600pc. This is in apparent contradiction with the
hypothesis of instantaneous local pollution proposed by \cite{KS86}.
Instead, this supports a picture where metals ejected in the current
burst of star formation escape into a hot halo hidden phase in the
halo, follow a long excursion while cooling and come back much later
into the central galactic region (or escape into the intergalactic
medium). This also implies that star
formation has been occurring previous to the current burst. Based on
different observational facts, we propose that the metals in IZw~18
are the result of a mild {\it continuous} star formation rate. The
generalization of this model to all gas clouds can account for the
scarcity of local galaxies with a metallicity lower than IZw~18, for
the increase with time of the metallicity of the most underabundant
DLA systems, and for the apparent absence of HI clouds without optical
counterparts. If starbursts appear as important episodes in the
history of galaxies, the low continuous star formation regime,
dominant during the quiescent inter-burst periods, cannot be
neglected.


\begin{acknowledgements}
This work is part of the PhD thesis of FL.
We thanks  P.~Petitjean, J.~Silk, M.~Fioc, R.~Terlevich, N. Prantzos,
F.~Combes, G.~Tenorio-Tagle and the referee R. Dufour for helpful
suggestion and discussions.  
\end{acknowledgements}

\bibliography{/home/legrand/LATEX/BIBLIOGRAPHIE/bibliographie}

\end{document}